\documentclass[aps,prd,reprint,twocolumn,superscriptaddress,nofootinbib]{revtex4-2}
\usepackage[utf8]{inputenc}   
\usepackage{amsmath,amssymb,amsfonts}
\usepackage{graphicx}
\usepackage{hyperref}
\usepackage{xcolor}
\usepackage[caption=false]{subfig}

\usepackage{amsmath,amsthm,amssymb}
\usepackage{graphicx}
\usepackage{dcolumn}
\usepackage{bm}
\usepackage{color}
\usepackage{epsfig}
\usepackage{multirow}
\usepackage{mathrsfs}
\usepackage{hyperref}
\usepackage{cleveref}
\usepackage{epstopdf}
\usepackage{autobreak}

\begin{document}

\title{Information paradox and island of covariant black holes in LQG}

\author{Yongbin Du}
	\affiliation{School of Physics and Astronomy, Sun Yat-sen University, Guangzhou, 510275, China}

    \author{Jia-Rui Sun}
   \thanks{Corresponding author}
    \email{sunjiarui@mail.sysu.edu.cn}
	\affiliation{School of Physics and Astronomy, Sun Yat-sen University, Guangzhou, 510275, China}
	
	\author{Xiangdong Zhang}
   \thanks{Corresponding author}
    \email{scxdzhang@scut.edu.cn}
	\affiliation{School of Physics and Optoelectronics, South China University of Technology, Guangzhou 510641, China}

\date{\today}

\begin{abstract}
We study information paradox of four dimensional covariant black holes inspired by loop quantum gravity (LQG) with two well motivated solutions. 
We first prepare the spacetime in the Hartle-Hawking state, compute the radiation entropy and recover a linear growth at late time.
When considering the mass loss and incorporating greybody factors, we show that for Metric~1 the LQG parameter $\zeta$ leaves temperature and Planckian factor of the spectrum unchanged but enhances the near-horizon barrier, leading to a faster evaporation rate as $M$ decreases. 
This behavior contrasts sharply with Metric~2, which has slow evaporation rate at small $M$ and admits a non-singular continuation suggestive of a remnant or a black-to-white-hole transition. 
We then apply the island prescription on the eternal background and find that quantum extremal surfaces exist in Metric~1 geometries; $\zeta$ gives a larger island
and the island suppresses the late time entropy growth, preserving unitarity. 
Our results highlight that covariance-respecting LQG black hole do not exhibit a universal late time behavior.
\end{abstract}

\maketitle

\section{introduction}


Black holes, emerging from general relativity, have continually challenged our understanding and unsettled beliefs about nature. An important manifestation lies in the fact that, at the fundamental mechanical level, the arrow of time is generally regarded as absent, and physical systems are, in principle, reversible. Yet the black holes render the situation different.
Classically, the event horizon functions as a one-way membrane that appears to break reversibility; quantum mechanically, however, it gives rise to Hawking radiation~\cite{Hawking:1974rv}, and restore the time arrow tension in a deeper form.
If a pure state collapses and fully evaporates, the late time state appears thermal, putting unitarity at risk~\cite{Hawking:1976ra}. 
This is the black hole information paradox: a quantum analogue of the entropy arrow, now phrased in terms of von Neumann entropy.
Since the paradox was first proposed, a vast literature has accumulated, crystallizing into a range of distinct viewpoints. Attitudes toward Bekenstein-Hawking entropy serve as a blunt but useful divider among the various approaches.
One line of thought interprets the Bekenstein-Hawking entropy as the maximum von Neumann entropy that a black hole, regarded as a quantum system, can possess \cite{Page:1993,tHooft:1993dmi,Susskind:1994vu}. In this view, if a collection of photons in a pure state collapses to form a black hole, then its horizon area is proportional to the system’s number of degrees of freedom. At late times in the evaporation process, Page’s theorem \cite{Page:1993} implies that the entanglement entropy of radiation particles cannot exceed the thermal entropy of the remaining black hole. Consequently, a unitary evaporation history necessarily yields a radiation entropy that follows the Page curve. This perspective gives a sharp expression to the holographic principle~\cite{tHooft:1993dmi,Susskind:1994vu,Maldacena:1997re,Gubser:1998bc,Witten:1998qj,Bousso:2002ju}.
A contrasting viewpoint holds that a black hole’s area entropy is merely the maximum entanglement entropy accessible to an exterior observer set by the surface states localized near the horizon. By contrast, a complete quantum description of the black hole may possess an enormous interior volume at late times, capable of accommodating an arbitrarily large number of degrees of freedom \cite{Rovelli:1996ti,Ashtekar:1997yu}. On this account, the information carried by the infalling apple is not lost: it may remain stored in the interior, persist in a remnant~\cite{Giddings:1992,Adler:2001,ChenOngYeom:2015}, erupt in a late time “fireworks”–like discharge~\cite{HaggardRovelli:2015,DeLorenzoPerez:2016,RovelliVidotto:2014}, or be hidden in vacuum–state degeneracies after the hole evaporates entirely~\cite{Ashtekar:2005cj,Callan:1992rs,Perez:2015,Perez:2017}. In any of these outcomes, reproducing the Page curve is not a necessary hallmark of unitary evaporation. The core significance of the second viewpoint is often manifested within the framework of loop quantum gravity (LQG).

Recently, an important development along the first line of thought is the proposal of island formula \cite{Penington:2019npb,Almheiri:2019psf,Almheiri:2019qdq}. Unlike the traditional approach of quantum field theory in curved spacetime, it computes the radiation entropy as a generalized entropy,
which contains the area term of the island boundary together with the finite part of entanglement entropy of the matter. It can be viewed as a generalization of the Ryu–Takayanagi (RT)/quantum extremal surface (QES) prescription of entanglement entropy~\cite{Ryu:2006bv,Hubeny:2007xt,Lewkowycz:2013nqa,Faulkner:2013ana,Engelhardt:2014gca}. To remove the backreaction of Hawking radiation on the geometry and thereby simplify the discussion of islands, Ref.~\cite{Almheiri:2019yqk} also proposes an ``eternal black hole'' version of the information paradox. In this setup, the black hole spacetime is glued to an external thermal bath such that the system radiates into the bath yet the black hole does not disappear (i.e., the background geometry is held fixed), providing a time-independent arena to analyze islands. In this background, an island typically appears outside the horizon and its volume grows in tandem with the expanding interior at late time, effectively offsetting the continued growth of the radiation entropy. It reproduces the expected Page curve, and the framework has since been extended to higher dimensional black holes and other gravity models~\cite{Hashimoto:2020cas,Almheiri:2019hni,Du:2022vvg,Tong:2023nvi,Yu:2021cgi,Yu:2021cgi,Ge:2025huo,Yu:2025tid}.
In this work, to examine whether the island formula can also be applied in LQG background, we construct an eternal black hole spacetime within LQG and compute both its Hawking radiation entropy and island entropy.

LQG is one of the promising way to quantize gravity with non-perturbative and background independent features. During the
past decades, the application of the quantization methods developed in LQG to black
hole models has drawn increasing attention, and many black
hole models with LQG corrections have been
proposed \cite{Black_Gambini_2008,Quantum_Ashtekar_2018,Loop_Zhang_2020,Effective_Kelly_2020,Loop_Zhang_2023,Effective_Lin_2024}. However, currently, most LQG black hole models suffer from implicit covariance issues \cite{Black_Gambini_2008,Quantum_Ashtekar_2018,Loop_Zhang_2020,Effective_Kelly_2020,Loop_Zhang_2023,Effective_Lin_2024}, where
physical predictions depend on coordinate choices and hence become unreliable. Recently, covariant
black hole solutions under Hamiltonian constraints have been found in \cite{Zhang:2024Cov,Black_Zhang_2025a,Black_Belfaqih_2025,Mass_Lin_2025} and shadows and innermost stable circular orbit have been studied in these backgrounds~\cite{Liu:2024soc,Du:2024ujg}. In particular, once backreaction is included, correction of the LQG effective metric from minimal covariance requirements tends to slow the late time evaporation rate \cite{Belfaqih:2025hawking}, which indicates that in the late stage of evaporation the evolution of the black hole is governed by new physics: it may either halt evaporation and leave behind a remnant, or undergo a transition into a white hole, thereby releasing the information hidden in its interior. In order to analyze Hawking emission on a different LQG black hole satisfying minimal covariance constraints and study the influence of quantum corrections on the black hole information paradox. In this paper we aim to investigate these issues in four dimensional covariant effective spacetimes.

The structure of this paper is as follows: In Section~2, we analyze the time evolution of radiation entropy in a fixed effective LQG geometry. In Section~3, we investigate the loop quantum gravity corrections to the greybody spectrum under the mass decay of the black hole. In Section~4, we search for possible islands in LQG black holes and evaluate the generalized entropy in the island phase. Finally, in Section~5, we give a brief discussion of our findings. Taken together, these results highlight the nontrivial role of LQG corrections in late-time black hole evaporation and suggest that the fate of black holes may strongly depend on the specific effective metric adopted.

\section{Radiation entropy on a Fixed Spacetime Background}

In this paper, we calculate the radiation entropy in the LQG geometry, which is the central outcome of~\cite{Zhang:2024Cov}. The guiding idea is that the classical diffeomorphism invariance, encoded in the constraint algebra of general relativity, must remain consistent once quantum modifications are introduced. To achieve this, the authors imposed two minimal conditions on the effective Hamiltonian constraint. Under these requirements, the canonical transformations generated by the Hamiltonian constraints correspond precisely to spacetime diffeomorphisms of an effective metric.
Two distinct effective metrics emerge in Boyer-Lindquist coordinates \((t, r, \theta, \phi)\) as
	\begin{eqnarray}
		ds^2 &=& -f[r] \, dt^2 + \frac{1}{f[r] h[r]} \, dr^2 + r^2 d\Omega^2
	\end{eqnarray}
	with metric 1:
	\begin{eqnarray}\label{solution1}
		f[r] &=& 1 - \frac{2M}{r} + \frac{\zeta^2 M^2}{r^2} \left(1 - \frac{2M}{r} \right)^2, \\
		h[r] &=& 1,
	\end{eqnarray}
	and metric 2:
	\begin{eqnarray}\label{solution2}
		f[r] &=& 1 - \frac{2M}{r}, \\
		h[r] &=& 1 + \frac{\zeta^2 M^2}{r^2} \left(1 - \frac{2M}{r} \right),
	\end{eqnarray}
	where \( M \) is the mass of the covariant black hole. The quantum parameter \( \zeta \) is defined as
    \begin{equation}
        \zeta = \frac{\sqrt{4\sqrt{3}\pi \gamma^3 \ell_p^2}}{M},
    \end{equation}
    with \( \gamma \) being the Barbero-Immirzi parameter and \( \ell_p \) the Planck length. It can be seen that when the black hole mass is large, the loop quantum correction parameter of the effective metric is very small. However, when the black hole mass approaches the Planck scale, the loop quantum behavior of the effective metric becomes distinctly pronounced.

	Metric~1 is similar in form to a Reissner-Nordström (RN) black hole with a small charge. The equation $f(r) = 0$ has two positive roots, given by $r_{+} = 2M$ and $r_{-} = M^{2}\zeta^{2}/\beta - \beta/3$, where $\beta^{3} = 3M^{3}\zeta^{2}(\sqrt{81 + 3\zeta^{2}} - 9)$. Similar to the characteristics of an RN black hole, these two roots correspond to the outer and inner horizons. Metric~2 also possesses two special zeros, but with different meanings~\cite{Belfaqih:2025hawking}. The first, determined by the root of $f(r)=0$ at $r=r_{h}$, is precisely the event horizon of the black hole, and it carries the same temperature as the Schwarzschild horizon.
The second arises from the condition $h(r)=0$, which defines an additional zero located at $r=r_{\Delta}$. It reads as
    \begin{equation}
r_{\Delta} = 3M \, \delta^{1/3} 
\frac{\left(1 + \sqrt{1+\delta}\right)^{2/3} - \delta^{1/3}}
{\left(1 + \sqrt{1+\delta}\right)^{1/3}},
\label{eq:xDelta}
\end{equation}
where $\delta = \Delta/27M^{2}$ and $\Delta=\zeta^2$. When taking $M \gg \sqrt{\Delta}$, we obtain $r_{\Delta}(M) \approx (2M\Delta)^{1/3}$. This radius is not an inner horizon; rather, it is a very small scale that represents a ``minimal area'' radius. In Metric~2 the spacetime does not terminate at $r=0$, but instead ``bounces'' or continues at $r=r_{\Delta}$, where the singularity is replaced by a nonsingular minimal radius. The corresponding Penrose diagram can thus be interpreted as a transition region that connects a black hole region to a white hole region.
 Both solutions are spherically symmetric, and in the limit $\zeta \to 0$ they reduce to the Schwarzschild spacetime.

Next, we discuss the black hole information paradox in this background. 
The original picture of the information paradox is as follows.  
In Minkowski spacetime, consider a collection of quanta in a pure state,
sent in from spatial infinity toward the center of spherical coordinates.  
When these quanta are compressed into a region whose total radius becomes smaller than the Schwarzschild radius of a system with the same mass, they undergo gravitational collapse and form a black hole.  
The black hole then emits blackbody radiation until it completely evaporates, and the spacetime returns from a curved geometry to a flat one.  In the end, the fine–grained entropy of the outgoing radiation is nonzero, whereas the initial state was pure with vanishing entropy, and this mismatch constitutes the information paradox. 

As mentioned in the first paragraph of the Introduction, the black hole information paradox discussed above can be reformulated in terms of the Page curve under the assumption that the black hole entropy does not exceed the area entropy \cite{Page:1993,Almheiri2021}. This assumption is known as the Central Dogma described in \cite{Almheiri2021}, which reads 

\textit{From the viewpoint of an external observer, a black hole can be described as a quantum system with \(\mathrm{Area}/(4G_N)\) degrees of freedom, which evolves unitarily in time.}\\
The assumption may be viewed as a manifestation of the holographic principle, at least from the perspective of an external observer of the black hole. 
In the late stage of black hole evaporation, unitarity requires the radiation entropy to equal the entropy of the black hole itself. Meanwhile, the Central Dogma asserts that the von Neumann entropy of the black hole cannot exceed its area entropy, which is getting smaller and smaller. This implies that the radiation entropy is bounded by the Bekenstein-Hawking area entropy and should therefore decrease monotonically with time. This behavior is in direct tension with the monotonic increase of the Hawking radiation entropy.
Thus, by assuming the Central Dogma, the non-unitarity of black hole evolution manifests itself most sharply at the moment when the radiation entropy exceeds the Bekenstein-Hawking area entropy, i.e. at the Page time.

\begin{figure}[!htb]
		\centering
			\includegraphics[width=0.4\textwidth]{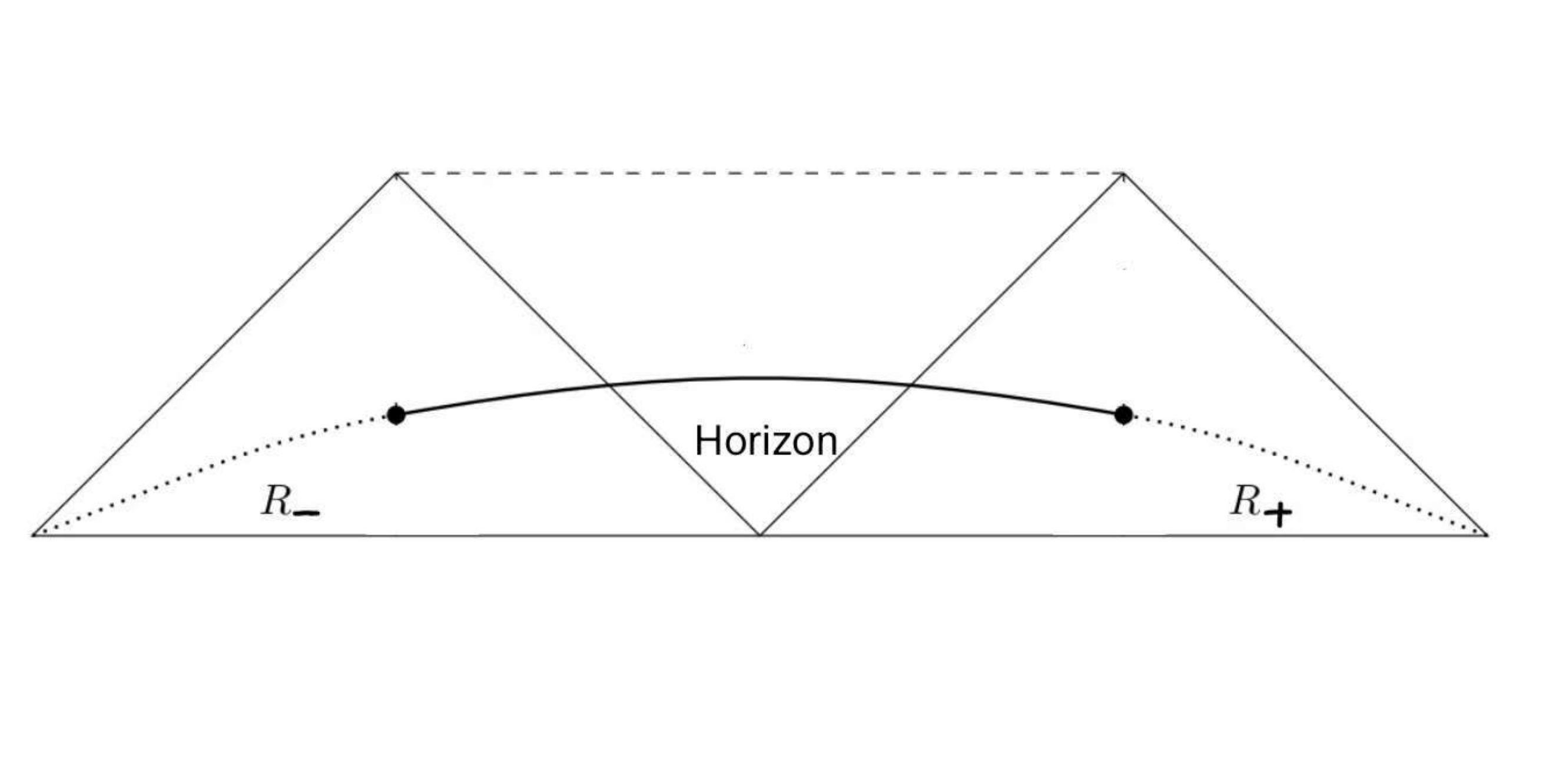}
		\caption{Penrose diagram of a spherically symmetric spacetime.  
The region used for the computation of radiation entropy is dotted and denoted by $R=R_{-}\cup R_{+}$.}
		\label{noisland}
	\end{figure}

This picture, drawn by Hawking and Page, is very intuitive, but its drawback is that any computation of the radiation entropy must take into account the backreaction of the collapsing matter on the spacetime geometry, which makes the problem technically difficult.  We will analyze this picture  in the next section.
A more practical strategy is to construct a situation in which the background geometry remains stationary: one introduces an external quantum state outside the black hole such that the two systems are in thermal equilibrium, so that the black hole no longer evaporates away but still exchanges quanta with the exterior.  
Under the assumption of Central Dogma, one may construct an approximate, quasi-static version of the black hole information paradox.
Following Ref.~\cite{Hashimoto:2020cas,Almheiri:2019hni,Du:2022vvg,Tong:2023nvi}, we likewise extend the spacetime to a two-sided geometry, construct an LQG eternal black hole, and place the Hartle-Hawking state on this background, as show in Fig.~\ref{noisland}.

On the Cauchy slice at $t = 0$, the Hartle-Hawking system outside the two-sided black hole is highly entangled between the left and right exteriors, while the combined state of the two sides is still pure.  
The two-sided black hole itself is also in a pure state, so that the entire Cauchy slice of the spacetime is globally pure.  
As time evolves, i.e.\ as we push the Cauchy slice forward along the time direction of the exterior observers, the black hole and the exterior Hartle-Hawking bath begin to couple and exchange quanta, thereby building up new entanglement between them.  
This entanglement keeps increasing, and the fine-grained entropy of
$R_{+}\cup R_{-}$, which is precisely the radiation entropy collected in the exterior system, grows accordingly.  
After the Page time, the radiation entropy exceeds the area entropy of an eternal black hole, while the black hole entropy— which should purify the radiation in order to preserve unitarity— is still bounded by the area according to the Central Dogma. This leads to a non-conservation of the total von Neumann entropy, namely the information paradox for eternal black hole.

We now proceed to compute the radiation entropy in the background of the LQG eternal black hole.
The renormalized entanglement entropy of matter fields of the exterior region $R=R_{-}\cup R_{+}$ can be computed as~\cite{Hashimoto:2020cas} 
\begin{equation}\label{finiteentropy}
S^{(\mathrm{finite})}_{\mathrm{matter}}(R) \;=\; -\, I\!\left(R_{+};\,R_{-}\right),
\end{equation}
where $I(R_{+};R_{-})$ denotes the mutual information between the disjoint exterior regions $R_{+}$ and $R_{-}$. For simplicity, we only focus on the late time regime of evaporation, which is sufficient to expose the information paradox. In this regime, $R_{-}$ and $R_{+}$ are separated by a distance much larger than the correlation length of the massive modes in the Kaluza–Klein tower associated with the spherical sector; hence only the $s$ wave contributes to the entanglement entropy. 
Therefore, the mutual information reduces to that of a two dimensional massless conformal field theory
\begin{equation}
I(A;B) = -\frac{c}{3}\,\log d(x,y),
\label{eq:mutual_info}
\end{equation}
where $c$ is the central charge and $d(x,y)$ denotes the conformal distance between $x$ and $y$, the boundaries of $A$ and $B$, respectively. 
Note that, because we employ the $s$–wave approximation, this formula is valid only when the boundaries of the two regions under consideration are strictly spherically symmetric.

\begin{figure}
    \centering
    \includegraphics[width=1\linewidth]{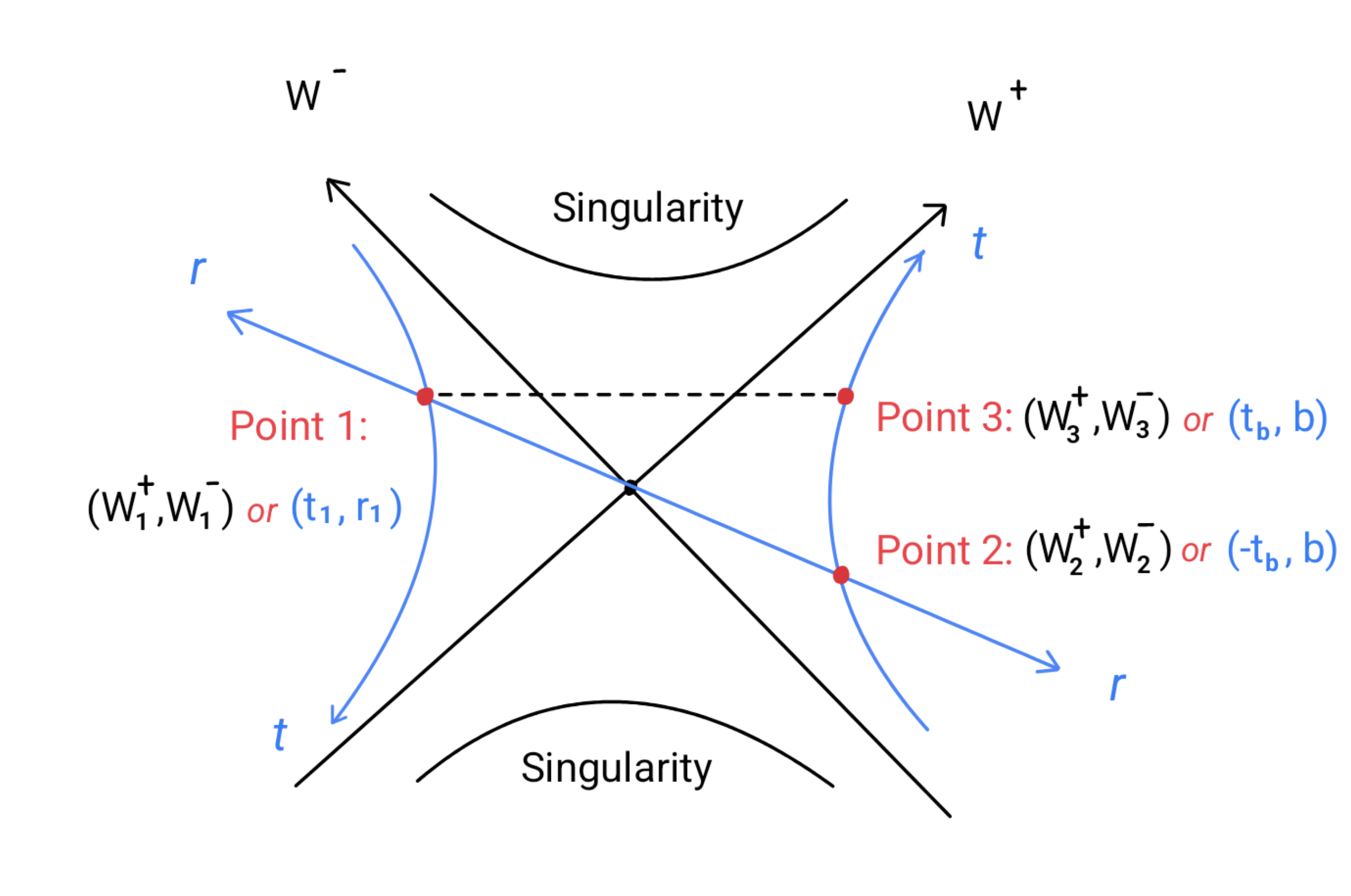}
    \caption{Kruskal coordinate $(W^{+},W^{-})$ and Schwarzschild coordinate $(t,r)$. Point 1 represent $b_{-}$, whose Schwarzschild coordinate is written as $(t_{1},r_{1})$ and Kruskal coordinate as $(W_{1}^{+},W_{1}^{-})$. Point 3 represent $b_{+}$, the coordinate of which is $(t_{b},b)$ or $(W_{3}^{+},W_{3}^{-})$.}
    \label{fig:placeholder}
\end{figure}

We first analyze the entanglement entropy in the Metric~1, given by eq. \eqref{solution1}. To recast the metric in Kruskal coordinates $(U,V)$, we first introduce the tortoise coordinate
\begin{equation}\label{tortoise}
    r_{*}^{(1)}(r) = \int^{r} \frac{dr'}{f(r')} \, .
\end{equation}
Then, adopting the Eddington–Finkelstein and Kruskal coordinates in the right exterior region,
\begin{align}
u &= t - r_{*}^{(1)}, \qquad 
v = t + r_{*}^{(1)}, \notag \\[6pt]
U &= - e^{-\kappa u}, \qquad 
V = e^{\kappa v}, \label{eq:kruskal}
\end{align}
where $\kappa=1/2f'(r_{h}) = 1/4M$, denotes the surface gravity of the black hole. The two-dimensional $(t,x)$ sector of the metric takes the form
\begin{equation}
    \left.ds^{2}\right|_{t,x} 
    = - \frac{f(r)}{\kappa^{2}} 
      e^{-2\kappa r_{*}^{(1)}(r)} \, dU dV \, .
\end{equation}
Including the spherical part, the full metric becomes
\begin{equation}
    ds^{2} = -\frac{1}{W^{2}} \, dU dV + r^{2} d\Omega^{2}, 
    \qquad 
    \frac{1}{W^{2}} = \frac{f(r)}{\kappa^{2}} \, e^{-2\kappa r_{*}^{(1)}(r)} \, .
\end{equation}
At late times, since the boundary points of the two intervals $R_{+}$ and $R_{-}$ are widely separated, the entanglement entropy is given by
\begin{equation}
    S_{\rm rad} = \frac{c}{3}\,\log d\!\left(b_{+}, b_{-}\right) .
\end{equation}
The coordinates of the boundary $b_{+}$ of $R_{+}$ are chosen as $(t,r)=(t_b,b)$. 
To express the entanglement entropy explicitly as a function of $t_{b}$ and $b$, we introduce the notation $W^{+}=V$ and $W^{-}=U$, which allows us to derive the coordinates of $b_{-}$ in a more convenient and transparent manner.
As illustrated in Fig.~\ref{fig:placeholder}, points~3 and~1 correspond to $b_{+}$ and $b_{-}$, respectively, and our goal is simply to express the coordinates of $b_{-}$ in terms of the real coordinates of $b_{+}$. We then introduce point 2, which is centrally symmetric to point 1.  In Kruskal coordinates its relation to point 3 is $W_{2}^{\pm} = W_{3}^{\mp}\, $.
 Taking into account that the coordinate transformations on the left and right patch are
    \begin{eqnarray}
    \text{ Left patch:}\,\,W^{\pm} = \mp\, e^{\,\kappa( t \pm r_*)};\\
    \text{Right patch:}\,\,W^{\pm} = \pm\, e^{\,\kappa (t \pm r_*)},
    \end{eqnarray}
    we find the relation in the right patch
    \begin{eqnarray}
W_{1}^{+} = -\, W_{3}^{-} &=& 
e^{-\,\kappa(t_{b}-r_{*}(b))}, \\[6pt]
W_{1}^{-} = -\, W_{3}^{+} &=& 
-\,e^{\kappa(t_{b}+r_{*}(b))}.
\end{eqnarray}
Substituting this into the left–patch coordinate transformation and using $\exp(i\pi) = -1$, we obtain $(t_{1}, r_{1}) = (-t_{b} + i\,4M\pi,\; b)\,=(-t_{b} + i\beta/2, b)$ with $\beta=2\pi/\kappa$.

Following a conformal map to Kruskal coordinates $(U,V)$ with Weyl factor $W$, the matter contribution becomes
\begin{equation}\label{entropy}
    S_{\rm rad} = \frac{c}{6}\,\log
    \frac{\big[U(b_{-})-U(b_{+})\big]\big[V(b_{+})-V(b_{-})\big]}
         {W(b_{+})\,W(b_{-})}\, .
\end{equation}
Substituting the coordinates into it, we obtain
\begin{align}
\frac{(U - U_+)(V_+ - V_-)}{W(b_+) W(b_-)}
= \frac{4 e^{2 \kappa r_*^b} \cosh^2(\kappa t_b)}{\kappa^2 e^{2 \kappa r_*^b}/f(b)}.
\end{align}
So we get
\begin{equation}\label{entropyformula}
    S_{\rm rad}(b,t_{b}) 
    = \frac{c}{6} \log \left[ \frac{4 f(b)}{\kappa^{2}} 
      \cosh^{2}(\kappa t_{b}) \right] .
\end{equation}
In the late–time limit, the entropy grows linearly in time,
\begin{equation}
S_{\rm rad}(t_b)\;\sim\;\frac{c}{6}\,\frac{t_b}{r_h},
\label{eq:late_time_linear_growth}
\end{equation}
Consequently, the radiation entropy inevitably surpasses the eternal black hole’s area entropy $2S_{\rm BH}$ in a finite time and never turns over. This reproduces, for LQG effective Metric~1 \eqref{solution1}, the eternal black hole version of the information paradox when taking Central Dogma assumption. In a fixed geometric background, LQG corrections enter the late time entropy only as an additive constant, and therefore do not affect the persistence of the paradox. 
Note that the discussion above is valid only when we have assumed that the Central Dogma holds.
This is because the radiation entropy exceeding the area entropy does not, by itself, directly imply an information paradox for an eternal black hole. 
Some studies on LQG have suggested that the Central Dogma, together with the holographic principle on which it is based, may require modification \cite{Rovelli:1996ti,Ashtekar:1997yu}. For instance, in the late stage of black hole evaporation, the number of interior degrees of freedom may significantly exceed the bound implied by the area entropy, which can be distinguished by the internal observer  \cite{Rovelli2025}. Also, as described at the end of the first paragraph of the Introduction, there exists a class of proposed resolutions to the black hole information paradox that do not satisfy the Central Dogma, yet are still able to realize unitary evolution to a certain extent
~\cite{Giddings:1992,Adler:2001,ChenOngYeom:2015,HaggardRovelli:2015,DeLorenzoPerez:2016,RovelliVidotto:2014,Ashtekar:2005cj,Callan:1992rs,Perez:2015,Perez:2017}.
Within such modified frameworks, the radiation of an eternal black hole no longer manifests a sharp paradox associated with non-unitary evolution.

\begin{figure}[!htb]
		\centering
			\includegraphics[width=0.35\textwidth]{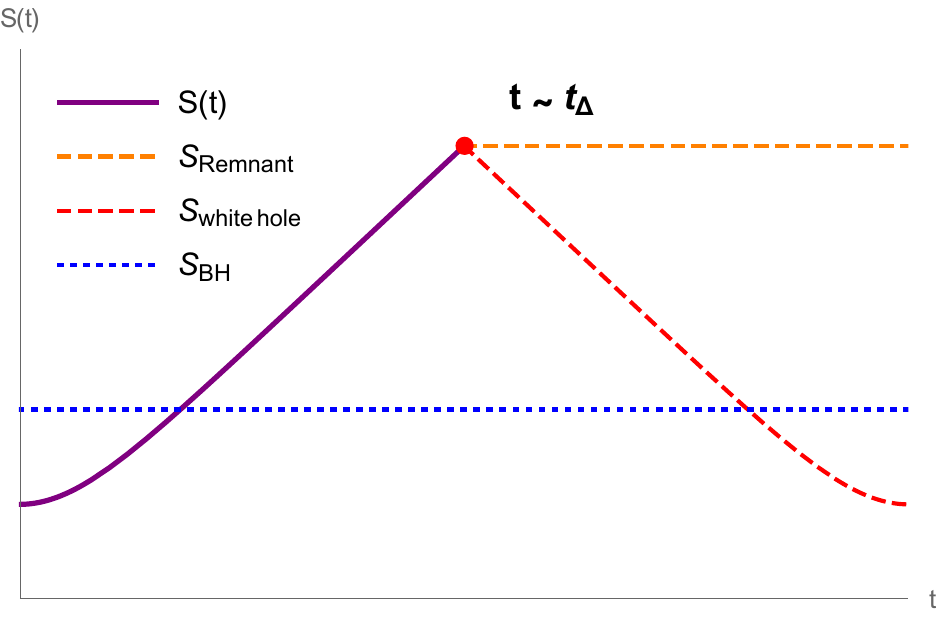}
		\caption{Possible late–time scenarios when the Cauchy surface intersects the minimal radius $r_\Delta$: entropy may saturate to a remnant value or decay in a white–hole transition.}
		\label{tdelta}
	\end{figure}

For LQG metric~2 \eqref{solution2} we likewise introduce the tortoise coordinate, defined by
\begin{equation}
    r^{(2)}_{*}(r) = \int^{r} \frac{dr'}{f(r') \sqrt{h(r')}} \, , 
\end{equation}
We still place the boundary of the observation region at radius $r=b$,
\begin{equation}
    b_{+}: (t,r) = (t_{b}, b), 
    \qquad 
    b_{-}: (t,r) = (-t_{b} + i\beta/2, b),
\end{equation}
with $\beta=2\pi/\kappa$ and surface gravity defined as $\kappa =  f'(r_{h}) \sqrt{h(r_{h})} /2  = 1/4M$. 
We also obtain the result \eqref{entropyformula}. 
However, the situation may change once we take into account the existence of a minimal area element at $r_{\Delta}$. For a dynamical black hole, the ultimate fate seen by an external observer remains uncertain. More puzzling, in the case of an eternal black hole, it is unclear whether the time-translated Cauchy surfaces will inevitably intersect the interior at $r=r_{\Delta}$. If, at some very late time $t_{\Delta}$, the Cauchy surface does intersect $r_{\Delta}$, this could signal the termination of the linear growth of radiation entropy. What follows might instead be a constant entropy associated with a remnant \cite{Giddings:1992,Adler:2001,ChenOngYeom:2015}, or a rapidly decaying entropy corresponding to a white-hole transition, as illustrated in Fig.~\ref{tdelta}.

The missing information may be preserved in the black hole or come back! This seems to suggest that, even in a fixed geometry where backreaction of the radiation is neglected but the singularity is removed, one can still formulate an apparently consistent unitary account of information, although the detailed mechanism remains open to debate.
  In such models, even if the Central Dogma is violated, at the semi-classical level, information may still be conserved. What is lacking is a concrete understanding of how to access or probe the information hidden inside the black hole.

\section{Radiation entropy of evaporating black hole}

As mentioned in the previous section, we now discuss the ``initial'' version of the LQG black hole information paradox. In this setup the black hole is single-sided and surrounded by vacuum, so there is nothing to prevent a Schwarzschild black hole from evaporating completely.
Why do we also need to analyze the information paradox in this version? The reason is that some discussions of LQG black holes have suggested that, when backreaction is taken into account and the metric of the black hole is allowed to evolve during evaporation, the LQG corrections to gravity may lead to results that deviate drastically from those of semiclassical calculations at late times.
These deviations may hint at a possible resolution of the black hole information paradox.
For example, in Ref.~\cite{Ashtekar:2008prl-arXiv}, a two dimensional semiclassical theory of quantum gravity with backreaction was analyzed using a bootstrap method, yielding a geometry that remains regular everywhere on the manifold and free of singularities. Within a mean-field approximation, it was further shown that the null infinity of the black hole spacetime coincides with that of Minkowski space, 
which provides preliminary evidence that taking backreaction into account may help in resolving the paradox. But it should also be noted that subsequent study \cite{Perez:2015} has shown that merely altering the mean field geometry near the singularity is not sufficient. The intrinsic spacetime discreteness, together with the associated degrees of freedom that may be invisible at the semiclassical level, appears to be an indispensable ingredient in resolving the information paradox.
In Ref.~\cite{Belfaqih:2025hawking}, the geometry under consideration coincides with LQG effective metric~2 \eqref{solution2} of the present work. There it was found that as the black hole mass decreases, the evaporation rate slows down significantly, also leaving room for a possible resolution of the paradox. These naturally raise the question: Does the LQG effective metric~1 \eqref{solution1} in our paper, once the mass loss of the black hole is taken into account, exhibit a slowdown of the evaporation rate—rather than reproducing the fixed background result of the previous section, which still displays the black hole information paradox? To address this, we will perform a preliminary analysis using the matching method (i.e. \cite{Harmark:2007jy}) to calculate the greybody factor of LQG geometry. 

Since the Hawking temperature of Metric 1 \eqref{solution1} coincides with that of Schwarzschild, the frequency–dependent particle spectrum is
\begin{equation}
\big\langle N_{\omega}\big\rangle
= \frac{\mathcal{T}_{\ell}(\omega)}{\exp\!\left(2\pi\,\omega\,\kappa^{-1}\right)-1}\,.
\end{equation}
where $\mathcal{T}_{\ell}(\omega)$ is the greybody (transmission) factor  and $\kappa=1/4M$.
As in Metric~2, the LQG parameter does not modify the Hawking temperature nor the Planckian factor of the spectrum. However, since the greybody factors are sensitive to the near–horizon effective potential, they can receive parameter–dependent corrections. For simplicity we consider a minimally coupled, massless scalar field. Its dynamics is governed by the Klein–Gordon equation
\begin{equation}
\Box \Phi_{\omega\ell} \;=\; \frac{1}{\sqrt{-g}}\,
\partial_{\mu}\!\left(\sqrt{-g}\,g^{\mu\nu}\partial_{\nu}\Phi_{\omega\ell}\right)=0\, 
\label{eq:KG_minimal_massless}
\end{equation}
on the static, spherically symmetric background. Decomposing $\Phi_{\omega\ell}(t,r,\Omega)=\sum_{\ell m} \psi_{\omega\ell}(r)\,Y_{\ell m}(\Omega)\,e^{i\omega t}/r$
and introducing the tortoise coordinate \eqref{tortoise}, Eq.~\eqref{eq:KG_minimal_massless} reduces to the one–dimensional Schrödinger form
\begin{equation}
\frac{d^{2}\psi_{\omega\ell}}{dr_*^{2}}
+\Big[\omega^{2}-V_{\ell}(r)\Big]\psi_{\omega\ell}=0,
\label{eq:radial_Schrodinger_general}
\end{equation}
where the effective potential $V_{\ell}(r)$ is written as 
\begin{equation}
    V_{l}(r)=f(r)\frac{f'(r)}{r}+f(r)\frac{l(l+1)}{r^2}
\end{equation}
We show its curve in Fig~\ref{effectivepotential} for different $\zeta$ when taking $l=0$.
\begin{figure}[!htb]
		\centering
			\includegraphics[width=0.35\textwidth]{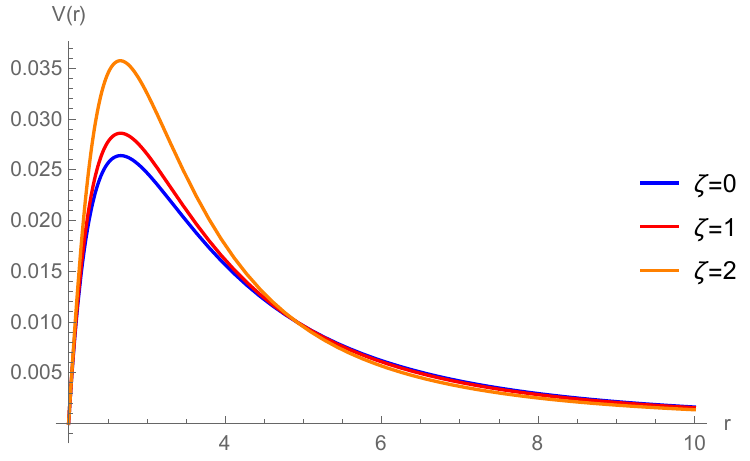}
		\caption{Effective potential $V(r)$ for different values of the LQG parameter $\zeta$ in the $l=0$ mode. A larger $\zeta$ increases the overall barrier height, affecting the greybody factor.}
		\label{effective potential}
	\end{figure}
Evidently, as $\zeta$ increases, the overall magnitude of the effective potential also grows, which in turn affects the value of the greybody factor. We resort to the low–frequency approximation, which means the typical wavelength of the radiation is much larger than the black hole size $r_{h}$, namely $r_{h}\,\omega \ll 1$ 
\cite{Belfaqih:2025hawking,Harmark:2007jy}. In this approximation, the dominant contribution to the greybody factor comes from the $l=0$ mode, and we shall therefore focus primarily on this sector. The equation of motion is then simplified as
\begin{equation}
\partial_{r}\!\left(r^{2} f(r)\, \partial_{r}\Phi_{\omega}\right)
+ \omega^{2}\,\frac{r^{2}}{f(r)}\,\Phi_{\omega}\;=\; 0 \, .
\end{equation}
To make the equation easier to analyze, one typically partitions the spacetime into three regions.
\begin{itemize}
  \item \textbf{Region I:} The near–horizon region, defined by $r \simeq r_{h}$ and $V(r) \ll \omega^{2}$.
  
  \item \textbf{Region II:} The intermediate region characterized by $V(r) \gg \omega^{2}$.
  
  \item \textbf{Region III:} The asymptotic region, defined by $r \gg r_{h}$.
\end{itemize}
Consider an incoming wave sent from spatial infinity (Region III) toward the black hole (Region I). Upon encountering the effective potential barrier, part of the wave is reflected back to infinity while the remainder is transmitted through the barrier and is ultimately absorbed at the horizon. Solving the Klein–Gordon equation in the three regions and using the conserved KG current density,
\begin{equation}
j \;=\; \frac{1}{2i}\!\left(\Phi_\omega^{*}\frac{d\Phi_\omega}{dr_{*}}
-\Phi_\omega\,\frac{d\Phi_\omega^{*}}{dr_{*}}\right),
\label{eq:KGcurrent}
\end{equation}
we can relate the incoming flux $J_{\text{in}}$, the reflected flux $J_{\text{out}}$, and the horizon (transmitted) flux $J_{\text{hor}}$:
\begin{equation}
J_{\text{in}} \;=\; J_{\text{out}} + J_{\text{hor}} .
\label{eq:flux_cons}
\end{equation}
The greybody factor for the $l=0$ mode is then defined as the transmission probability,
\begin{equation}
\mathcal{T}_{0}(\omega)\;\equiv\;\frac{J_{\text{hor}}}{J_{\text{in}}}\,.
\label{eq:greybody_def}
\end{equation}

For Region I, the solution has a simple form
\begin{equation}
\Phi_{\omega}(r_{*}) \;=\; A_{\mathrm{I}}\, e^{i\omega r_{*}}\,,
\end{equation}
so its associated flux can be expressed as
\begin{equation}
J_{\mathrm{hor}}
= \frac{1}{2i}\!\left(\Phi_\omega^{*}\frac{d\Phi_\omega}{dr_{*}}
-\Phi_\omega\,\frac{d\Phi_\omega^{*}}{dr_{*}}\right) A_{H}
= 4\pi r_{h}^{2} \omega\,|A_{\mathrm{I}}|^{2}\,,
\end{equation}
where $A_{H}$ is the area of the horizon. In the low–frequency limit, the solution in Region I can be expanded as  
\begin{equation}
\Phi_{\omega}(r) \;=\; A_{\mathrm{I}}
\left[
1 \,+\, i\,\omega r_{*}
\right] .
\end{equation}
To exhibit the effect of the LQG parameter, we retain terms up to order $\zeta$ in the near–horizon tortoise coordinate. The solution can then be expanded as  
\begin{equation}
\Phi_{\omega}(r) \;=\; A_{\mathrm{I}}
\left[
1 \,+\, i\,\frac{\omega}{2\kappa}
\left(
\log\!\left(\frac{r-r_{h}}{r_{h}}\right) + i\,\omega\,\alpha \,r_{h}
\right)
\right] ,
\end{equation}
where $\alpha=-\zeta^2/4$ encodes the $\zeta$–dependent corrections specific to the effective metric under consideration.

Then we move to Region II. 
Its most general solution can be written as  
\begin{equation}
\Phi_{\omega}(r) \;=\; A_{\mathrm{II}} \,+\, B_{\mathrm{II}}\, G(r) \, ,
\end{equation}
with  
\begin{equation}
G(r) \;=\; \int_{\infty}^{r} \frac{dr'}{r^2 f(r')},  .
\end{equation}
Near the horizon $r \simeq r_{h}$, this expression simplifies to  
\begin{equation}
G(r) \;\simeq\; \frac{1}{2 r_{h}^{\,2}\,\kappa} 
\log\!\left(\frac{r-r_{h}}{r_{h}}\right).
\end{equation}
In the overlap region the two solutions must match, yields the relations 
\begin{equation}
A_{\mathrm{II}} \;=\; A_{\mathrm{I}}[1+i\,\omega\,\alpha\,r_{h}], 
\qquad 
B_{\mathrm{II}} \;=\; i\,\omega\,r_{h}^{\,2}\,A_{\mathrm{I}} \, .
\end{equation}
In the asymptotic region far from the horizon one finds $G(r)$ take the form of $-1/r$. This allows us to connect the solution with that of Region~III in the overlap domain.

\begin{figure}[!htb]
  \centering
  \subfloat[Evaporation rates $\mathrm{d}M/\mathrm{d}t$ of Metric~1 for $\zeta=0$ and $\zeta=\sqrt{3}$, shown as functions of the black hole mass $M$.]{%
    \includegraphics[width=0.35\textwidth]{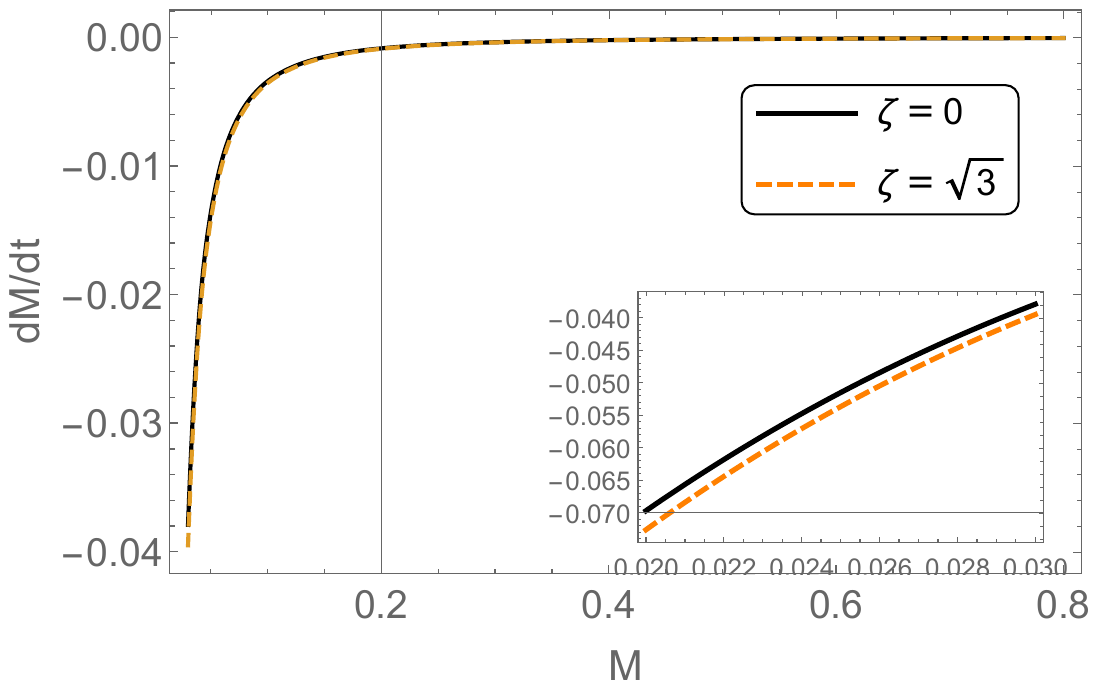}%
  }\hspace{1em}
  \subfloat[Difference in evaporation rates, $\Delta_{rad} (\mathrm{d}M/\mathrm{d}t)$, between $\zeta=\sqrt{3}$ and $\zeta=0$, highlighting the impact of loop–quantum corrections.]{%
    \includegraphics[width=0.35\textwidth]{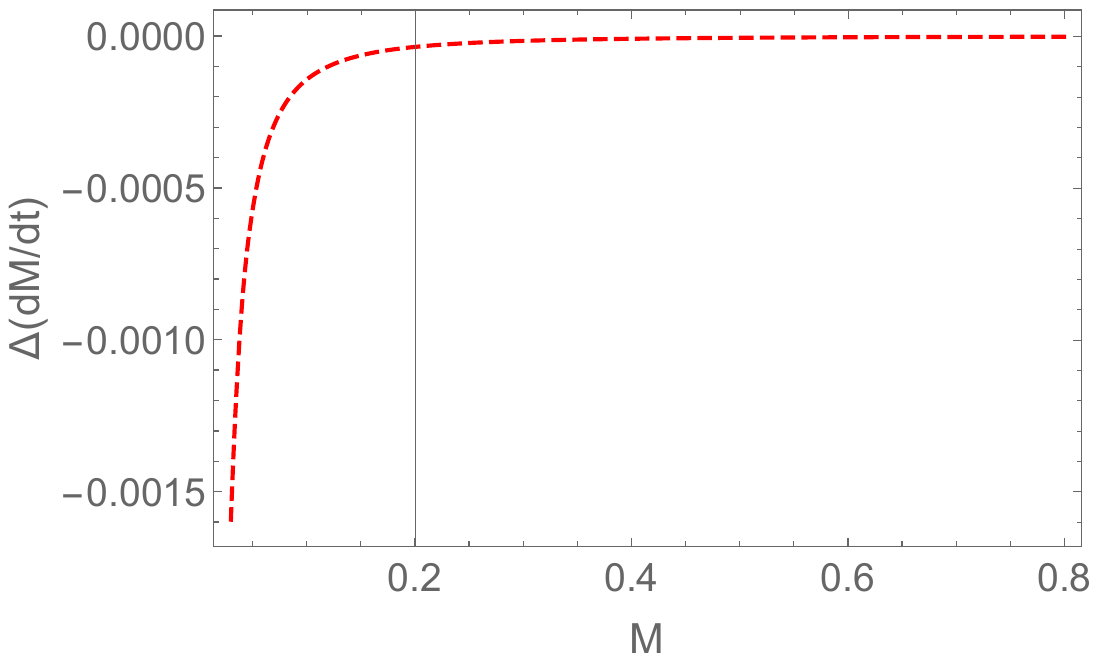}%
  }
  \caption{Evaporation rates for the two cases.}
  \label{evaporationrate}
\end{figure}

In Region~III the general solution takes the form  
\begin{equation}
\Phi_{\omega}(r) 
= \rho^{-1/2}
\left[
C_{1}\, H^{(1)}_{1/2}(\rho)
+ C_{2}\, H^{(2)}_{1/2}(\rho)
\right].
\end{equation}
where $\rho=\omega r$ and $H^{(1)}_{1/2}(\rho)$ are Hankel functions. For $\rho \ll 1$, the solution can be expanded as  
\begin{equation}
\Phi_{\omega}(r) \;\simeq\;
\frac{C_{1}+C_{2}}{\sqrt{2\pi}/2}\,
- i\frac{\sqrt{2\pi}}{\pi\omega r}\,(C_{1}-C_{2}) .
\end{equation}
Here $C_{1}$ corresponds to the amplitude of the incoming wave in the asymptotic region. Comparing the two solutions in Region II and Region III, we get the expression of greybody factor as
\begin{equation}
    \mathcal{T}_{0}(\omega)
    =\frac{4\pi r_{h}^2 \omega |A_{\mathrm{I}}|^{2}}{8|C_{\mathrm{1}}|^{2}/\omega}
    =\frac{4r_{h}^2\omega^2}{(1-\omega^2r_{h}^2)^2+(4-\alpha^2)\omega^2r_{h}^2}
\end{equation}
In the limit $\alpha \to 0$, we recover the Schwarzschild black hole result \cite{Harmark:2007jy}. 
By substituting the greybody factor into the spectrum formula and integrating over all frequencies, the evaporation rate of the black hole is obtained as  
\begin{equation}
\frac{dM}{dt}
= -\,\frac{1}{2\pi}\int_{0}^{\infty} d\omega \,
\frac{\omega\,\mathcal{T}_{0}(\omega)}
{\exp\!\left(2\pi\omega/\kappa\right)-1} \, .
\end{equation}
We plot the evaporation rate as a funtion of $M$ for $\zeta=\sqrt{3}$ and $\zeta=0$ in Fig~\ref{evaporationrate}. To make it clear to see the difference between the rates, we define
\[\Delta_{rate}(\frac{dM}{dt}):=\left.\frac{dM}{dt}\right|_{\zeta=\sqrt{3}}
-\left.\frac{dM}{dt}\right|_{\zeta=0},
\]
and show its curve as a function of the black hole mass in Fig.~\ref{evaporationrate}.

In Ref.~\cite{Belfaqih:2025hawking}, the authors also studied the evaporation rate of black holes in an LQG background, using a spacetime metric whose form coincides exactly with our ``Metric~2,'' namely Eq.~\eqref{solution2}.  Their result shows that a nonzero LQG parameter slows down the late-time evaporation and may even halt the process altogether, in sharp contrast with Hawking's prediction that the
evaporation accelerates toward the final stage.
In the present section, however, we analyze a different black hole geometry, ``Metric~1,'' given in Eq.~\eqref{solution1}, which is another physically reasonable effective metric consistent with the covariant requirements of LQG.  Our calculation reveals the opposite behavior:
for this metric the late time evaporation does not slow down; instead,
the LQG parameter $\zeta$ further enhances the late time evaporation rate, even compared to the standard Hawking result.
This indicates that effective LQG metrics satisfying the covariance constraints do not exhibit a universal late time behavior: the $\zeta$ corrections in the second effective metric, Eq.~\eqref{solution2}, tend to slow the evaporation, suggesting an endpoint with a remnant or a preparation for a black–to–white–hole transition. In contrast, the trend in the Fig.~\ref{evaporationrate} for LQG covariant black hole spacetime, Eq.~\eqref{solution1}, shows no such sign. In this spacetime the black hole information paradox persists, calling for a more complete quantum gravity description to resolve it.

    \section{Island in LQG geometry}
The analysis in the previous two sections shows that, for the Metric~1 in Eq.~\eqref{solution1} considered in this paper, the information paradox under Central Dogma assumption has no signal that it can be removed by LQG corrections, regardless of whether backreaction is taken into account.
Therefore, in the final part of this work, for simplicity, we return to the two-sided eternal black hole background and investigate whether an island can be found in this setting, how such an island differs from the case without LQG corrections, and what the appearance of an island implies for LQG itself.

The conceptual origin of the island formula can be traced back to the RT prescription~\cite{Ryu:2006bv} and its quantum extension, the QES formula~\cite{Engelhardt:2014gca}, which includes all orders of quantum corrections.  The QES formula takes the form
\begin{equation}
S(A)=\min\,\Bigg\{\; \mathrm{ext}\left[
\frac{\mathrm{Area}(\gamma_{A})}{4\,G_{N}}
+ S_{\mathrm{matter}}(\Sigma)
\right]\Bigg\},
\label{eq:island_main}
\end{equation}
where $\gamma_{A}$ is a codimension-two bulk minimal surface homologous to the boundary subregion
$A$, and it is determined by varying it in the bulk so as to
extremize $S(A)$ and then selecting the minimal extremum. $\Sigma$ denotes the bulk region bounded by $\gamma_{A}$ and
$A$.  The matter fields living in $\Sigma$ contribute an entropy
$S_{\mathrm{matter}}(\Sigma)$, which can be evaluated using
semiclassical methods. Both the RT and QES prescriptions were originally developed within the framework of the
holographic duality, where the goal is to calculate the
entanglement entropy of strongly coupled quantum fields from the dual bulk gravitational theory.

\begin{figure}
    \centering
    \includegraphics[width=0.95\linewidth]{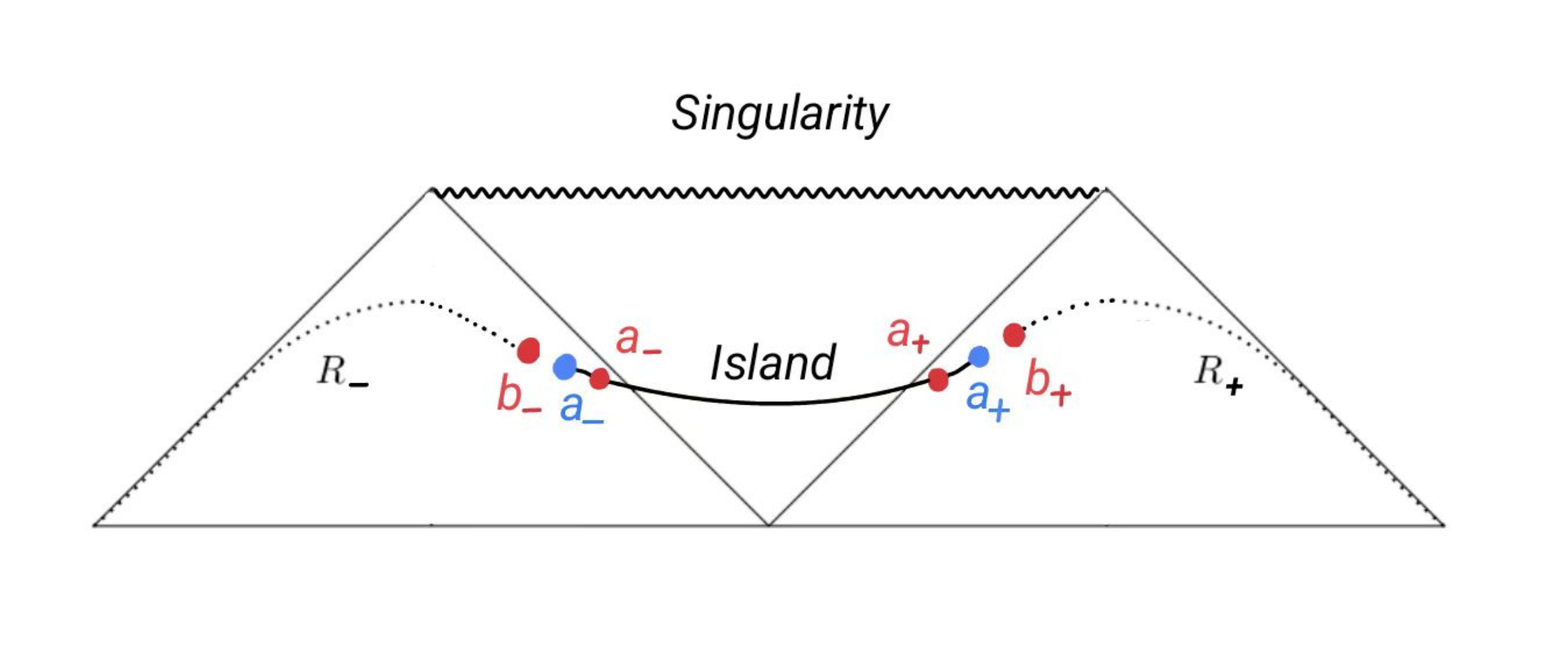}
    \caption{Island in Schwarzschild and LQG metric 1 in Eq.~ \eqref{solution1}. For nonzero $\zeta$, the island exist, and the radial location of the boundary $a_{\pm}$ of it (marked in blue points) is larger than the Schwarzschild case (marked in red points).}
    \label{Schwarzschild_island}
\end{figure}

Can this prescription then guide us in computing the entanglement entropy in a curved spacetime with a black hole? The answer is yes, but one needs to engineer an appropriate spacetime setup. For example, in Ref.~\cite{Almheiri:2019yqk}, the authors glued an AdS$_2$ spacetime endowed with a JT dilaton field to two flat Minkowski regions, and introduced CFT thermal baths in the Minkowski parts so that they are in thermal equilibrium with the AdS$_2$ black hole.  This construction provided a toy model of an eternal black hole. A key feature of the model is that the AdS$_2$ region is dual to a quantum mechanical (QM) system on the boundary.
Now assume that the two baths and the AdS$_2$ black hole are initially
in pure states, and that for $t>0$ the black hole is allowed to exchange energy with the baths. Their interaction causes the entanglement entropy of the baths to grow, and this entropy is precisely identified with the radiation entropy of the black hole.
Therefore, when computing the entanglement entropy of the thermal baths, it is equivalent to computing the entanglement entropy of the QM systems themselves. The latter can be further evaluated using the $AdS_{2}$ QES prescription. So the entropy of the radiation naturally takes the structural form of a QES formula as 
 \begin{equation}
S(R)=\min\,\Bigg\{\; \mathrm{ext}\left[
\frac{\mathrm{Area}(\partial I)}{4\,G_{N}^{}}
+ S_{\mathrm{matter}}(R\cup I)
\right]\Bigg\}.
\label{island}
\end{equation}
which is called the island formula~\cite{Penington:2019npb,Almheiri:2019psf,Almheiri:2019qdq}, where $R=R_{-}\cup R_{+}$ denotes the radiation region and $I$ represents the island region whose boundary is exactly QES. This is the island formula. We illustrate the island in Fig.~\ref{Schwarzschild_island}.  In this setup, it successfully reproduces the Page curve and thus preserves the unitarity of the evolution. It was shown in~\cite{Almheiri:2019qdq} that by including new saddle points of wormholes in gravitational path integral over replica geometries, the island formula is not restricted to the holographic dual setting, but is likely the correct prescription for computing black hole entanglement entropy in generic curved spacetimes. Thus the present work studies black hole radiation entropy in LQG using this formula. 

To apply the island formula, the divergent part of entanglement entropy of matter fields can be absorbed into the area term in eq.~\eqref{island}, which leads to a renormalized version of the island formula~\cite{Hashimoto:2020cas}
\begin{equation}
S(R)=\min\,\Bigg\{\;\mathrm{ext}\left[
\frac{\mathrm{Area}(\partial I)}{4\,G_{N}^{(r)}}
+ S_{\mathrm{matter}}^{(\mathrm{finite})}(R\cup I)
\right]\Bigg\},
\label{eq:island_main}
\end{equation}
with the renormalized Newton coupling defined by
\begin{equation}
\frac{1}{4\,G_{N}^{(r)}} \;=\; \frac{1}{4\,G_{N}} \;+\; \frac{1}{\varepsilon^{2}},
\label{eq:GN_ren}
\end{equation}
where $\varepsilon$ is the short distance cutoff. At late time, due to large distance between $R_{-}$ and $R_{+}$, the matter sector can be calculated through mutual information at each side
\begin{equation}
S_{\mathrm{matter}}^{(\mathrm{finite})}(R\cup I)
\;=\; -\,2\,I(R_{+};\,I).
\label{eq:matter_mutual_info}
\end{equation}
However, since the separation between $R_{+}$ and the island can be arbitrarily small, the $s$ wave no longer dominates the mutual information thus the two dimensional CFT entropy expression is inapplicable. Instead we use \cite{Casini:2009dt,Casini:2009}
\begin{equation}
I(A;B) \;=\; k\, c\, \frac{\mathrm{Area}}{L^{2}}\,,
\end{equation}
where $c$ is the central charge, $L$ is the proper distance between the boundaries of $A$ and $B$, and $k$ is a constant.

Still choosing the coordinate of the boundary of $R$ as $(t,r)=(t_b,b)$ for $b_{+}$ and $(t,r)=(-t_b+i\beta/2,b)$ for $b_{-}$. We only consider the case $b-r_{h} \ll r_{h}$, where the radiation region lies very close to the black hole horizon. In this regime, the entanglement entropy can be evaluated using Eq.~\eqref{eq:island_main} and Eq.~\eqref{eq:matter_mutual_info}, and takes the form
\begin{equation}
    S_{\text{gen}}(a) 
    = \frac{2\pi a^{2}}{G_{N}}
      - 2 k\, c \, \frac{4\pi b^{2}}{L^{2}} \, .
\end{equation}
where $L$ is defined as
\begin{equation}
    L = \int_{a}^{b} \frac{dr}{\sqrt{f(r)}} \, .
\end{equation}
When both $a$ and $b$ are located near the horizon ($a<b$), we set 
$r = r_{h} + \epsilon$. In this limit, $L$ can be rewritten in terms of $\epsilon$ as
\begin{equation}
    L =\large \int_{\epsilon_{a}}^{\epsilon_{b}} \frac{d\epsilon}{\sqrt{\frac{\epsilon}{r_{h}+\epsilon}+\frac{\zeta^2 M^2 \epsilon^2}{(r_{h}+\epsilon)^4}}} \, .
\end{equation}
Expanding $L$ in terms of $\epsilon$, we obtain
\begin{align*}
    L &= \int_{\epsilon_{a}}^{\epsilon_{b}} d\epsilon  \large(
    \sqrt{\frac{r_{h}}{\epsilon}}
    + \frac{r_{h}^{2} - (M\zeta)^{2}}{2\,r_{h}^{5/2}}\,\sqrt{\epsilon}
    + \mathcal{O}(\epsilon^{3/2}) \, )\\[3pt]
    &\simeq \int_{\epsilon_{a}}^{\epsilon_{b}} d\epsilon(c_{1}\epsilon^{-1/2}+c_{2}\epsilon^{1/2}).
\end{align*}
with $c_{1}=\sqrt{r_{h}}$ and $c_{2}=(r_{h}^{2} - M^2\zeta^{2})/2\,r_{h}^{5/2}$.
From the above expression, it is clear that when $a$ and $b$ are extremely close to the horizon, the loop quantum gravity corrections to the entanglement entropy become negligibly small. In this limit, one recovers the same quantum extremal surface and entanglement entropy as in the Schwarzschild spacetime. However, in order to exhibit the qualitative effects of the loop quantum gravity parameter, we retain the expansion of $L$ up to order $\epsilon^{1/2}$.
In the near-horizon limit, the proper distance can be approximated as
\begin{align*}
    L \;&\simeq\; 2 c_{1} \left( \sqrt{\,b-r_{h}\,} - \sqrt{\,a-r_{h}\,} \right) \\
    &+ \frac{2}{3}c_{2} ((b-r_{h})^{3/2}-(a-r_{h})^{3/2}).
\end{align*}
We introduce the notations
\begin{equation}
    x = \sqrt{\frac{a-r_{h}}{r_{h}}}, 
    \qquad 
    \beta = \sqrt{\frac{b-r_{h}}{r_{h}}} \, .
\end{equation}
Then the proper distance can be written as
\begin{equation}
    L = 2 c_{1}^{2} (\beta - x) 
      + \frac{2}{3} c_{2} c_{1}^{3} \left( \beta^{3} - x^{3} \right) .
\end{equation}

In the near-horizon limit $b \simeq r_{h}$ and $x \ll 1$, 
the extremization condition $\partial S/\partial a=0$ is equivalent to $\partial S/\partial x=0$, which yields
\begin{equation}
    x \Bigg[ (\beta - x) 
    + \frac{1}{3} c_{1} c_{2} \beta^{3} \Bigg]^{3}
    = \frac{G_{N} c k}{2 c_{1}^{4}} \, .
\end{equation}
In the semiclassical approximation, the right-hand side of the equation becomes extremely small. Therefore, the left-hand side can be expanded in powers of $x$ to solve the equation. Eventually, we find that the solution for $x$ is
\begin{equation}
    x=\frac{G_{N} c k}{2 c_{1}^{4} \beta_{0}^{3}}
\end{equation}
where $\beta_{0}=\beta+c_{1}c_{2}\beta^{3}/3$. When $\zeta=0$, $\beta_{0}$ reduces to the Schwarzschild result. Finally, we obtain the value of $a$
\begin{equation}
a = r_h \left( 1 + x^2 \right) 
= r_h + \frac{(G_{N} c k)^2}{4 r_{h} \beta_0^6}
\end{equation}
which is outside the horizon, and the correponding entanglement entropy 
\begin{equation}
S_{\text{gen}} 
\simeq \frac{2 \pi r_h^2}{G_N} 
- \frac{2 \pi k c b^2}{r_h^2 \beta_0^2}.
\end{equation}
Obviously, the entropy is a constant, and the first term is Bekenstein-Hawking entroy. For nonzero values of $\zeta$, $c_{2}$ becomes smaller, which shifts the location of the quantum extremal surface $a$ to larger values, which is shown in Fig.~\ref{Schwarzschild_island}. Consequently, in the effective spacetime with loop quantum gravity corrections, the island extends further. Evidently, an island can be identified in the LQG geometry; the quantum parameter $\zeta$ shifts the location of the island boundary $\partial I$. 
Physically, a larger island means that a larger portion of the black hole interior is encoded in the radiation degrees of freedom. In our setup, the LQG corrections systematically increase the radial location of the island boundary compared to the Schwarzschild black hole case. This indicates that LQG effects modify the entanglement wedge associated with Hawking radiation, while leaving the qualitative Page–curve behavior intact. We believe that this provides a concrete way in which LQG corrections affect the information–theoretic structure of black hole evaporation.


     \section{Discussion}

Our analysis has pursued three complementary threads. First, on a fixed LQG-inspired effective geometry we tracked the time dependence of the Hawking radiation entropy and confirmed a linear growth at late time, reproducing the eternal–black–hole version of the information paradox under Central Dogma. Second, allowing the mass to decay and incorporating greybody factors, we quantified how loop quantum corrections enter the emission rate. For Metric~1 in Eq.~\eqref{solution1}, the parameter $\zeta$ leaves the Planckian factor and temperature unchanged but modifies the near horizon potential, slightly increasing the evaporation rate as $M$ decreases. Third, we examined the island prescription on the eternal background and found that a QES can exist even in the LQG setting; the role of $\zeta$ there is essentially to displace the island boundary rather than to nullify it. Taken together, these results indicate that covariance-respecting LQG effective metrics need not share a universal late time behavior and that the fate of information depends sensitively on the specific effective metric.

We want to stress again that the black hole information paradox formulated in terms of the Page curve is discussed under the assumption of the Central Dogma, which is conjectured to be hold in quantum gravity but has not been proofed so far.
Thus, some works in LQG do not endorse the information paradox as formulated by the Page curve \cite{Rovelli2025}.
In our work, we 
 find metric~2 admits a viable mechanism explaining why information is not lost even if the Central Dogma is violated. 
By contrast, for the geometric structure of metric~1, at present it remains unclear whether there exists a transparent mechanism within loop quantum gravity that guarantees unitary evolution (such as hidden discrete spacetime degrees of freedom), especially at the semiclassical level. Clarifying this issue constitutes an important direction for future work.
This is precisely our motivation for analyzing its non-unitary behavior and the emergence of islands within the framework of the Central Dogma.


A central qualitative contrast emerges between the two covariant geometry. In Metric~2 in Eq.~\eqref{solution2}, the appearance of an interior minimal radius $r_{\Delta}$ replaces the classical singularity by a transition surface and tends to slow down the evaporation as $M\to \mathcal{O}(\ell_{P})$, leaving room for a remnant or a black–to–white–hole transition. In Metric~1, by contrast, we found no sign of such a turnover: the entropy continues to grow on a fixed background, and when mass loss is included the evaporation rate increases as $M$ shrinks. If nature selects an effective geometry akin to Metric~1, then islands become indispensable to preserve unitarity at least on semi-classical level; but if it selects something closer to Metric~2, the geometry itself may regulate the late time dynamics.

Our island analysis on the eternal background indicated that QES surfaces may not in conflict with the LQG modifications considered here. Operationally, $\zeta$ acts as a geometric control parameter that shifts $\partial I$ but does not obstruct extremization of $S_{\rm gen}$. This reinforced the idea that the island mechanism is robust beyond holographic toy models and two dimensional dilaton gravity, extending to four dimensional, covariance–constrained effective geometries. The open questions are whether such QES persist and remain dominant on time–dependent LQG backgrounds, and why LQG modification will give a larger island to encode the radiation degrees of freedom.


In future work, we hope to investigate more deeply whether there exists a $t_{\Delta}$ in Metric~1—that is, what kind of radiation entropy would be observed by an exterior observer of a permanent black hole. Besides, once dynamical evolution is taken into account, the ultimate fate of LQG black holes remains a persistent and important question. Following the various indications calculated in this paper, we may possess more tools to probe into the very depths of quantum gravity itself.

\begin{acknowledgments}
J.R.S. was supported by the National Natural Science Foundation of China (No.~12475069) and Guangdong Basic and Applied Basic Research Foundation (No.~2025A1515011321). X. Zhang was supported by the National Natural Science Foundation of China (No.~12275087). 
\end{acknowledgments}

\bibliographystyle{unsrt}

\begin{thebibliography}{10}
	



\bibitem{Hawking:1974rv}
S.~W.~Hawking,
``Black hole explosions,''
Nature \textbf{248} (1974), 30-31

\bibitem{Hawking:1976ra}
S.~W.~Hawking,
``Breakdown of predictability in gravitational collapse,''
Commun. Math. Phys. \textbf{43} (1976), 199--220.



\bibitem{Page:1993}
D.~N.~Page,
``Information in black hole radiation,''
Phys. Rev. Lett. \textbf{71} (1993), 3743--3746
[arXiv:hep-th/9306083]


\bibitem{tHooft:1993dmi}
G.~'t Hooft,
``Dimensional reduction in quantum gravity,''
Conf. Proc. C \textbf{930308} (1993), 284-296
[arXiv:gr-qc/9310026 [gr-qc]].

\bibitem{Susskind:1994vu}
L.~Susskind,
``The World as a hologram,''
J. Math. Phys. \textbf{36} (1995), 6377-6396
[arXiv:hep-th/9409089 [hep-th]].


\bibitem{Maldacena:1997re}
J.~M.~Maldacena,
``The Large $N$ limit of superconformal field theories and supergravity,''
Adv. Theor. Math. Phys. \textbf{2} (1998), 231--252
[Int. J. Theor. Phys. \textbf{38} (1999), 1113--1133]
[arXiv:hep-th/9711200 [hep-th]].

\bibitem{Gubser:1998bc}
S.~S.~Gubser, I.~R.~Klebanov and A.~M.~Polyakov,
``Gauge theory correlators from noncritical string theory,''
Phys. Lett. B \textbf{428} (1998), 105--114
[arXiv:hep-th/9802109 [hep-th]].

\bibitem{Witten:1998qj}
E.~Witten,
``Anti de Sitter space and holography,''
Adv. Theor. Math. Phys. \textbf{2} (1998), 253--291
[arXiv:hep-th/9802150 [hep-th]].



\bibitem{Bousso:2002ju}
R.~Bousso,
``The Holographic principle,''
Rev. Mod. Phys. \textbf{74} (2002), 825--874
[arXiv:hep-th/0203101 [hep-th]].

\bibitem{Rovelli:1996ti}
C.~Rovelli,
``Black hole entropy from loop quantum gravity,''
Phys.\ Rev.\ Lett.\  \textbf{77} (1996), 3288-3291
[arXiv:gr-qc/9603063 [gr-qc]].

\bibitem{Ashtekar:1997yu}
A.~Ashtekar, J.~Baez, A.~Corichi and K.~Krasnov,
``Quantum geometry and black hole entropy,''
Phys.\ Rev.\ Lett.\  \textbf{80} (1998), 904-907
[arXiv:gr-qc/9710007 [gr-qc]].

\bibitem{Giddings:1992}
S.~B.~Giddings,
``Black holes and massive remnants,''
Phys.\ Rev.\ D \textbf{46} (1992), 1347--1352.

\bibitem{Adler:2001}
R.~J.~Adler, P.~Chen and D.~I.~Santiago,
``The Generalized uncertainty principle and black hole remnants,''
Gen.\ Rel.\ Grav.\ \textbf{33} (2001), 2101--2108
[arXiv:gr-qc/0106080].

\bibitem{ChenOngYeom:2015}
P.~Chen, Y.~C.~Ong and D.~h.~Yeom,
``Black hole remnants and the information loss paradox,''
Phys.\ Rept.\ \textbf{603} (2015), 1--45
[arXiv:1412.8366 [gr-qc]].



\bibitem{HaggardRovelli:2015}
H.~M.~Haggard and C.~Rovelli,
``Black hole fireworks: Quantum-gravity effects outside the horizon spark black to white hole tunneling,''
Phys.\ Rev.\ D \textbf{92} (2015), 104020
[arXiv:1407.0989 [gr-qc]].

\bibitem{DeLorenzoPerez:2016}
T.~De~Lorenzo and A.~Perez,
``Improved black hole fireworks: Asymmetric black-hole-to-white-hole tunneling scenario,''
Phys.\ Rev.\ D \textbf{93} (2016), 124018
[arXiv:1512.04566 [gr-qc]].

\bibitem{RovelliVidotto:2014}
C.~Rovelli and F.~Vidotto,
``Planck stars,''
Int.\ J.\ Mod.\ Phys.\ D \textbf{23} (2014), 1442026
[arXiv:1401.6562 [gr-qc]].




\bibitem{Ashtekar:2005cj}
A.~Ashtekar and M.~Bojowald,
``Black hole evaporation: a paradigm,''
Class.\ Quant.\ Grav.\ \textbf{22} (2005) 3349–3362.

\bibitem{Callan:1992rs}
C.~G.~Callan, S.~B.~Giddings, J.~A.~Harvey and A.~Strominger,
``Evanescent black holes,''
Phys.\ Rev.\ D \textbf{45} (1992) R1005–R1009.

\bibitem{Perez:2015}
A.~Perez,
``No firewalls in quantum gravity: the role of discreteness of quantum geometry in resolving the information loss paradox,''
Class.\ Quant.\ Grav.\ \textbf{32} (2015) 084001.


\bibitem{Perez:2017}
A.~Perez,
``Black holes in loop quantum gravity,''
Rept.\ Prog.\ Phys.\ \textbf{80} (2017), 126901
[arXiv:1703.09149 [gr-qc]].



\bibitem{Penington:2019npb}
G.~Penington,
``Entanglement wedge reconstruction and the information paradox,''
JHEP \textbf{09} (2020), 002
[arXiv:1905.08255 [hep-th]].

\bibitem{Almheiri:2019psf}
A.~Almheiri, N.~Engelhardt, D.~Marolf and H.~Maxfield,
``The entropy of bulk quantum fields and the entanglement wedge of an evaporating black hole,''
JHEP \textbf{12} (2019), 063
doi:10.1007/JHEP12(2019)063
[arXiv:1905.08762 [hep-th]].

\bibitem{Almheiri:2019qdq}
A.~Almheiri, T.~Hartman, J.~Maldacena, E.~Shaghoulian and A.~Tajdini,
``Replica wormholes and the entropy of Hawking radiation,''
JHEP \textbf{05} (2020), 013
[arXiv:1911.12333 [hep-th]].



\bibitem{Ryu:2006bv}
S.~Ryu and T.~Takayanagi,
``Holographic derivation of entanglement entropy from AdS/CFT,''
Phys.\ Rev.\ Lett.\ \textbf{96} (2006), 181602
[arXiv:hep-th/0603001].

\bibitem{Hubeny:2007xt}
V.~E.~Hubeny, M.~Rangamani and T.~Takayanagi,
``A covariant holographic entanglement entropy proposal,''
JHEP \textbf{07} (2007), 062
[arXiv:0705.0016 [hep-th]].

\bibitem{Lewkowycz:2013nqa}
A.~Lewkowycz and J.~Maldacena,
``Generalized gravitational entropy,''
JHEP \textbf{08} (2013), 090
[arXiv:1304.4926 [hep-th]].

\bibitem{Faulkner:2013ana}
T.~Faulkner, A.~Lewkowycz and J.~Maldacena,
``Quantum corrections to holographic entanglement entropy,''
JHEP \textbf{11} (2013), 074
[arXiv:1307.2892 [hep-th]].


\bibitem{Engelhardt:2014gca}
N.~Engelhardt and A.~C.~Wall,
``Quantum extremal surfaces: holographic entanglement entropy beyond the classical regime,''
JHEP \textbf{01} (2015), 073
[arXiv:1408.3203 [hep-th]].


\bibitem{Almheiri:2019yqk}
A.~Almheiri, R.~Mahajan and J.~Maldacena,
``Islands outside the horizon,''
[arXiv:1910.11077 [hep-th]].


\bibitem{Hashimoto:2020cas}
K.~Hashimoto, N.~Iizuka and Y.~Matsuo,
``Islands in Schwarzschild black holes,''
JHEP \textbf{06} (2020), 085
[arXiv:2004.05863 [hep-th]].

\bibitem{Almheiri:2019hni}
A.~Almheiri, R.~Mahajan and J.~E.~Santos,
``Entanglement islands in higher dimensions,''
JHEP \textbf{08} (2020), 094
[arXiv:1911.09666 [hep-th]].

\bibitem{Du:2022vvg}
D.~H.~Du, W.~C.~Gan, F.~W.~Shu and J.~R.~Sun,
``Unitary constraints on semiclassical Schwarzschild black holes in the presence of island,''
Phys. Rev. D \textbf{107}, no.2, 026005 (2023)
[arXiv:2206.10339 [hep-th]].

\bibitem{Tong:2023nvi}
C.~W.~Tong, D.~H.~Du and J.~R.~Sun,
``Island of Reissner-Nordstr{\"o}m anti{\textendash}de Sitter black holes in the large D limit,''
Phys. Rev. D \textbf{109}, no.10, 104053 (2024)
[arXiv:2306.06682 [hep-th]].

\bibitem{Yu:2021cgi}
M.~H.~Yu and X.~H.~Ge,
``Islands and Page curves in charged dilaton black holes,''
Eur. Phys. J. C \textbf{82} (2022) no.1, 14
[arXiv:2107.03031 [hep-th]].

\bibitem{Ge:2025huo}
X.~H.~Ge and C.~Zhang,
``On the Replica Problem in Supersymmetric SYK Models,''
[arXiv:2508.09611 [hep-th]].

\bibitem{Yu:2025tid}
M.~H.~Yu, S.~Y.~Lin and X.~H.~Ge,
``Replica Wormholes, Modular Entropy, and Capacity of Entanglement in JT Gravity,''
[arXiv:2501.11474 [hep-th]].


		\bibitem{Quantum_Ashtekar_2018}
		A. Ashtekar, J. Olmedo, and P. Singh, Quantum Transfiguration of Kruskal Black Holes, Phys. Rev. Lett. \textbf{121}, 241301 (2018).
		
		\bibitem{Loop_Zhang_2020}
		C. Zhang, Y. Ma, S. Song, and X. Zhang, Loop quantum Schwarzschild interior and black hole remnant, Phys. Rev. D \textbf{102}, 041502 (2020).
		
		\bibitem{Effective_Kelly_2020}
		J. G. Kelly, R. Santacruz, and E. Wilson-Ewing, Effective loop quantum gravity framework for vacuum spherically symmetric spacetimes, Phys. Rev. D \textbf{102}, 106024 (2020).
		
		\bibitem{Loop_Zhang_2023}
		X. Zhang, Loop Quantum Black Hole, Universe 9, 7 (2023).
		
		\bibitem{Effective_Lin_2024}
		J. Lin and X. Zhang, Effective four-dimensional loop quantum black hole with a cosmological constant, Phys. Rev. D \textbf{110}, 026002 (2024).
		
		
		\bibitem{Black_Gambini_2008}
		R. Gambini and J. Pullin, Black Holes in Loop Quantum Gravity: The Complete Space-Time, Phys. Rev. Lett. \textbf{101}, 161301 (2008).
		
		
\bibitem{Zhang:2024Cov}
		C. Zhang, J. Lewandowski, Y. Ma, and J. Yang, Black holes and covariance in effective quantum gravity, Phys. Rev. D \textbf{111}, L081504 (2025).
		
		\bibitem{Black_Zhang_2025a}
		C. Zhang, J. Lewandowski, Y. Ma, and J. Yang, Black Holes and Covariance in Effective Quantum Gravity: A Solution without Cauchy Horizons, arXiv:2412.02487.
		
		
		\bibitem{Black_Belfaqih_2025}
		I. H. Belfaqih, M. Bojowald, S. Brahma, and E. I. Duque, Black holes in effective loop quantum gravity: Covariant holonomy modifications, Phys. Rev. D \textbf{112}, (2025).
		
		
		\bibitem{Mass_Lin_2025}
		J. Lin, X. Zhang, and M. Bravo-Gaete, Mass inflation and strong cosmic censorship conjecture in the covariant quantum black hole, Phys. Rev. D \textbf{111}, 106025 (2025).


\bibitem{Liu:2024soc}
W.~Liu, D.~Wu and J.~Wang,
``Light rings and shadows of static black holes in effective quantum gravity,''
Phys. Lett. B \textbf{858} (2024), 139052
[arXiv:2408.05569 [gr-qc]].

\bibitem{Du:2024ujg}
Y.~Du, Y.~Liu and X.~Zhang,
``Spinning particle dynamics and the innermost stable circular orbit in covariant loop quantum gravity,''
JCAP \textbf{05} (2025), 045
[arXiv:2411.13316 [gr-qc]].


\bibitem{Belfaqih:2025hawking}
I.~H.~Belfaqih, M.~Bojowald, S.~Brahma and E.~I.~Duque,
``Hawking evaporation and the fate of black holes in loop quantum gravity,''
[arXiv:2504.11998 [gr-qc]].


\bibitem{Harmark:2007jy}
T.~Harmark, J.~Natario and R.~Schiappa,
Adv. Theor. Math. Phys. \textbf{14} (2010)
[arXiv:0708.0017 [hep-th]].


\bibitem{Ashtekar:2008prl-arXiv}
A.~Ashtekar, V.~Taveras and M.~Varadarajan,
``Information is Not Lost in the Evaporation of 2D Black Holes,''
Phys. Rev. L \textbf{100}, 211302 (2008).

\bibitem{Casini:2009dt}
H.~Casini and M.~Huerta,
``Remarks on the entanglement entropy for disconnected regions,''
JHEP {\bf 03} (2009) 048
[arXiv:0812.1773 [hep-th]].


\bibitem{Casini:2009}
H.~Casini and M.~Huerta,
``Entanglement entropy in free quantum field theory,''
J.\ Phys.\ A {\bf 42} (2009) 504007
[arXiv:0905.2562].

\bibitem{Almheiri2021}
A.~Almheiri, T.~Hartman, J.~Maldacena, E.~Shaghoulian, and A.~Tajdini,
``The entropy of Hawking radiation,''
\emph{Rev.\ Mod.\ Phys.}\ \textbf{93}, 035002 (2021).

\bibitem{Rovelli2025}
C.~Rovelli,
``Black Holes Have More States than Those Defined by the Bekenstein--Hawking Entropy: A Simple Argument,''
\emph{Universe} \textbf{11}(1), 6 (2025)

        
	\end{thebibliography}

\end{document}